# A semi-analytical model for the propagation of a relativistic jet in a magnetized medium

Leonardo García-García[1]★, Diego López-Cámara[2], Davide Lazzati[3]
[1]*Instituto de Astronomía, Universidad Nacional Autónoma de México*
[2]*Investigador por México, CONAHCyT – Universidad Nacional Autónoma de México, Instituto de Astronomía, AP 70-264, CDMX 04510, México*
[3]*Department of Physics, Oregon State University, 301 Weniger Hall, Corvallis, OR 97331, U.S.A.*

**ABSTRACT**
The merger of two magnetized compact objects, such as neutron stars, forms a compact object which may launch a relativistic and collimated jet. Numerical simulations of the process show that a dense and highly magnetized medium surrounds the system. This study presents a semi-analytical model that models the effects that a static magnetized medium with a tangled field produces in relativistic, collimated, and non-magnetized jets. The model is a first approximation that addresses the magnetic field present in the medium and is based on pressure equilibrium principles between the jet, cocoon, and external medium. A fraction of the ambient medium field is allowed to be entrained in the cocoon. We find that the jet and cocoon properties may be affected by high magnetic fields ($\gtrsim 10^{15}$ G) and mixing. The evolution of the system may vary up to $\sim 10\%$ (compared to the non-magnetized case). Low-mixing may produce a slower-broader jet with a broader and more energetic cocoon would be produced. On the other hand, high-mixing could produce a faster-narrower jet with a narrow and less-energetic cocoon. Two-dimensional hydrodynamical simulations are used to validate the model and to constrain the mixing parameter. Although the magnetic field and mixing have a limited effect, our semi-analytic model captures the general trend consistent with numerical results. For high magnetization, the results were found to be more consistent with the low mixing case in our semi-analytic model.

## 1 INTRODUCTION

Short Gamma-ray Bursts (SGRBs) are one of the brightest transient events in the universe (Gehrels & Mészáros 2012). They are intense and brief flashes of gamma-ray radiation with isotropic energies in the range of $\sim 10^{50} - 10^{52}$ erg (Gehrels et al. 2009) with a duration of $\lesssim 2$ s (Kouveliotou et al. 1993). Despite their brevity, SGRBs are of great interest to astrophysicists as they offer valuable insights into the physics of extreme cosmic events. Their study contributes to the understanding of the dynamic and violent processes produced in the universe.

SGRBs originate from the merger of compact binary systems, mostly neutron star-neutron star (NS-NS) mergers (Abbott et al. 2017a). The evolution of relativistic and collimated jets through the ejecta launched during the NS-NS will, in turn, produce the prompt gamma emission (Abbott et al. 2017a,b; Goldstein et al. 2017; Hallinan et al. 2017; Margutti et al. 2017; Troja et al. 2017; Lazzati et al. 2017; Ruiz, Paschalidis, & Shapiro 2017; Lazzati et al. 2018; Mooley et al. 2018; Ghirlanda et al. 2019).

The SGRB-jets have velocities $\Gamma \sim 20-10^3$ (Ghirlanda et al. 2018) and isotropic luminosities $L_{iso} \sim 10^{50} - 10^{53}$ erg s$^{-1}$ (Ghirlanda et al. 2009). Their opening angles $\theta_j \sim 5° - 25°$ (Fong et al. 2015; Rouco Escorial et al. 2023), thus, the jets have luminosities of order $L_j \sim 10^{47} - 10^{51}$ erg s$^{-1}$. As the jet evolves through the medium, it deflects the material in its front producing a hot-pressurized bubble commonly known as the "cocoon" (Murguia-Berthier et al. 2014; Nagakura et al. 2014; Gottlieb et al. 2018; Lazzati, López-Cámara, & Morsony 2018; Hamidani et al. 2020; Sharan Salafia & Ghirlanda 2022). The cocoon then expands sideways and helps to collimate and shape the jet (Urrutia et al. 2021). Meanwhile, observations and GRMHD simulations of the merger of two NSs have found that the medium through which the jets evolve has a total mass of order $\sim 0.01 - 0.05 M_\odot$ (Cowperthwaite et al. 2017; Drout et al. 2017; Kasliwal et al. 2017; Smartt et al. 2017; Tanaka et al. 2017; Ruiz, Tsokaros, & Shapiro 2021), densities $\sim 10^8 - 10^{14}$ g cm$^{-3}$, and magnetic field strengths $\sim 10^{12} - 10^{16}$ G (Ciolfi et al. 2017; Combi & Siegel 2023; Ruiz, Tsokaros, & Shapiro 2021).

The merger of two magnetized compact objects, such as neutron stars, forms a compact object which may launch a relativistic and collimated jet. Numerical simulations of the process show that a dense and highly magnetized medium ($\sim 10^{15}$ G) surrounds the system (Paschalidis, Ruiz, & Shapiro 2015; Ruiz, Shapiro, & Tsokaros 2018, 2021; Sun et al. 2022, e.g.). In fact, it has been shown that the B field may be amplified up to intensities or order $\sim 10^{16}$ G due to turbulence and instabilities (Ciolfi et al. 2017; Aguilera-Miret et al. 2020; Mösta et al. 2020, e.g.).

To understand the evolution and morphology of SGRBs analytical models (Begelman & Cioffi 1989; Lazzati & Begelman 2005; Bromberg et al. 2011; Murguia-Berthier et al. 2014; Lazzati & Perna 2019; Hamidani & Ioka 2021; Gottlieb & Nakar 2022; Hamidani et al. 2024, e.g.) and numerical hydrodynamical simulations (De Colle et al. 2018a,b; Gottlieb, Nakar, & Piran 2018; Gottlieb et al. 2018; Lazzati et al. 2020; Hamidani et al. 2020; Gottlieb et al. 2021; Gottlieb & Globus 2021; Urrutia et al. 2021; Urrutia, De Colle, & López-Cámara 2023; De Colle et al. 2022, e.g.) have been conducted. Such studies have established an important role of the medium in the shaping of the jet. However, the role of the magnetic field of the medium remains to be fully understood.

---

★ E-mail: lgarcia@astro.unam.mx

García-García, López-Cámara, & Lazzati (2023) (GG23) carried out a set of two-dimensional relativistic hydrodynamical simulations to study the effect of embedding a poloidal magnetic field in the ejecta. They found that the presence of magnetized media in the cocoon reduces the number of recollimation shocks in the jet, and thus, the internal shocks have reduced efficiency. Furthermore, the magnetization of the cocoon increases when the jet-cocoon passes through a highly magnetized medium. Pavan et al. (2023) on the other hand, studied the dynamics of a magnetized jet drilling through a magnetized environment with general relativistic magneto-hydrodynamical simulations. They found that the medium collimates the jet more effectively and prevents the formation of turbulence. Such studies set the ground for future investigations to study the impact of the magnetization of the medium on the global evolution of SGRBs.

Even though analytic models present the global evolution and may not give information about the small-scale effects, these also present advantages over numerical simulations. For example, resolution effects, initial condition effects, small-scale domains, short integration times, inaccuracies in turbulence modeling, high-computational power demand are not present in the analytic framework. The semi-analytical model presented in this study builds upon the study of Lazzati & Perna (2019). In the latter, the analytic model was constructed based on three pressure balances to describe the structure of the jet-cocoon resulting from its interaction with an expanding and non-magnetized medium. Thus, the effects that a magnetized medium produces in the evolution of a relativistic jet remain to be understood.

In this study, for the first time, we construct a semi-analytical model to investigate the impact of a magnetized static medium on the evolution of a relativistic and non-magnetized jet and its cocoon. Our study is limited to a medium embedded with a tangled magnetic field. In addition, we compare the results from the semi-analytical model with those from a set of two-dimensional (2D) relativistic hydrodynamical (RHD) simulations. This paper is organized as follows. In Section 2 we describe the semi-analytical model. The discussion of the magnetic effects and the mixing is presented in Section 3. Section 4 compares the semi-analytic model with numerical simulations. Finally, the summary and conclusions are presented in Section 5.

## 2 SEMI-ANALYTICAL MODEL FOR A JET-COCOON IMMERSED IN A MAGNETIZED MEDIUM

The evolution of a relativistic and non-magnetized top-hat jet through a static medium depends on the jet luminosity, its opening angle, and the injection Lorentz factor, as well as on the density, pressure, and magnetization of the medium. The jet head propagates through the ejecta driving a bow shock that inflates a cocoon of hot material that enshrouds the jet. The cocoon exerts pressure on the jet, causing it to narrow. Additionally, the cocoon is trapped as well inside the ejecta and drives a shock into it.

Lazzati & Perna (2019) (hereafter LP19) study the jet and cocoon morphology through the pressure balances between the jet head and a moving medium, the jet, and the cocoon, and the cocoon and the medium. To understand the general effects that a magnetized medium produces in the relativistic and non-magnetized jet evolution we build upon LP19 by adding the magnetic pressure in the pressure balances that dominate the evolution of the system. Although the medium in SGRBs is expanding (with $\sim 0.2c$), for a fast-moving jet head (i.e., $c \gg 0.2c$) one can roughly approximate that the medium be considered as static. Moreover, we also consider that the medium has a tangled, fully randomized magnetic field, so the magnetic pressure is akin to an additional thermodynamic pressure component. We use the variable $\lambda$ to parameterize the fraction of the magnetized medium that is entrained in the cocoon.

The main goal of the model is to obtain the jet head velocity ($v_{jh}$), the opening angle of the jet ($\theta_j$), the opening angle of the cocoon ($\theta_c$), and the energy of the cocoon ($E_c$) as a function of the properties of the engine and of the medium (where the magnetic pressure and the mixing of the magnetized medium in the cocoon will be taken into account).

While the cocoon's shape is more accurately represented by an ellipsoid (Hamidani & Ioka 2021), we simplify our analysis by approximating it as a cylinder with a volume $V = r^3 \Omega_c$. Meanwhile for the jet, apart from the $v_{jh}$ and $\theta_j$, the jet is characterized by its luminosity ($L_j$). The medium is modeled as an ideal gas with density ($\rho_m$) and pressure ($P_m$) (with constant temperature and adiabatic index of 5/3). Based on the results of (Ciolfi et al. 2017) the density profile is a smoothed function of the result of the model in which two equal mass NSs (with a soft EoS) collide and in which the jet is launched 45 ms after the merger:

$$\rho_m(r) = \rho_0 \left(\frac{r_0}{r}\right)^n, \quad (1)$$

where $\rho_0$ is the density at the radius $r = r_0$, $n$ indicates the rate at which the density changes (e.g. for a constant medium $n = 0$, for a wind-like medium $n = 2$, and for the results obtained by Ciolfi et al. (2017) $n = 3$) [1]. We also assume that the magnetic field of the medium follows a power law distribution (as well as its correspondent magnetic pressure, $P_{m,B} \propto B^2$):

$$B_m(r) = B_{0,m} \left(\frac{r_0}{r}\right)^q, \quad (2)$$

where $B_{0,m}$ is the magnetic field at $r_0$ and $q$ is the rate at which the magnetic field varies (where for example, for a wind like medium, $q = 1$ implies that the magnetic field is frozen into the medium while $q = 3$ implies a constant ratio between $P_m$ and $P_{m,B}$). For the B frozen in the density case with $n = 3$, we have $q = 1.5$.

Consistently with LP19, the pressure balances that determine the global evolution of the jet and cocoon through the medium are: a) the balance at the head of the jet and the medium, where the ram pressure of the jet ($P_{j,ram}$) is balanced at the contact discontinuity by the ram pressure of the medium on the jet ($P_{mj,ram}$) plus the magnetic pressure of the medium ($P_{m,B} = \frac{B_m^2}{8\pi}$); b) the balance at the transition between the jet and the cocoon, where the jet internal pressure ($P_j$) is balanced by the cocoon thermal pressure $P_c$ and a fraction ($\lambda$, where 0 represents no mixing and 1 maximum mixing) of the medium's magnetic pressure which has entrained into the cocoon; and c) the balance at the transition between the cocoon and medium, where the $P_c$ and same fraction of the entrained magnetic pressure balance with the ram pressure of the medium on the cocoon

---

[1] Note that the adopted profile with $n = 3$ cannot extend to infinity since this would cause the ejecta mass to be infinity. An outer cutoff needs to be present to make this density profile physically acceptable. The density profile would also have to flatten in the core of the star to avoid a divergence at $r = 0$.

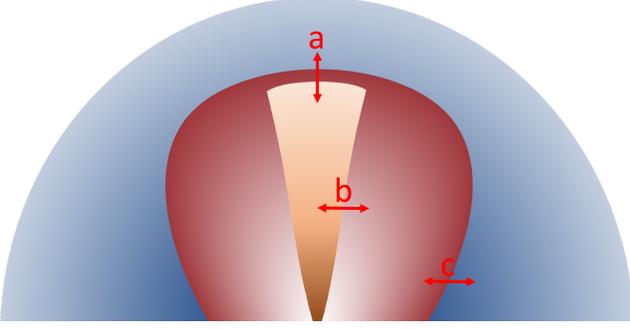

**Figure 1.** Scheme of the model. The jet, cocoon, and magnetized medium are shown in orange, red, and blue, respectively. The pressure balances a (see Equation 3), b (Equation 4), and c (Equation 5) used in our model are also indicated.

($P_{mc,ram}$) plus the $P_{m,B}$. These balances are next listed (and shown in Figure 1):

$$P_{j,ram} = P_{mj,ram} + P_{m,B}, \quad (3)$$

$$P_j = P_c + \lambda P_{m,B}, \quad (4)$$

$$P_c + \lambda P_{m,B} = P_{mc,ram} + P_{m,B}. \quad (5)$$

The ram pressure of the wind material on the cocoon is given by $P_{mc,ram} = \rho_m v_\perp^2 + P_{m,B}$, where $v_\perp^2 = \Omega_c r^2/\pi t^2$ is the squared velocity of the shock front driven by the cocoon into the ejecta at the radius $r = r(t)$, and $\Omega_c$ is the solid angle of the cocoon.

From Equation 3 assuming that the jet is relativistic ($v_j \sim c$), that the ram pressure of the jet and medium is greater than their correspondent internal pressures (i.e. a strong shock), that the enthalpy of the jet is pressure dominated, and changing the reference system to that which moves with the working surface of the jet, then the $v_{jh}$ is (see the Appendix A for further details):

$$v_{jh} = c \left\{ \frac{1 - \left[1 - \left(1 - \frac{\rho_m c^3 r^2 \Omega_j}{L_j}\right)\left(1 - \frac{P_{m,B} r^2 \Omega_j c}{L_j}\right)\right]^{1/2}}{1 - \frac{\rho_m c^3 r^2 \Omega_j}{L_j}} \right\}, \quad (6)$$

where $\Omega_j$ is the solid angle of the jet.

The opening angle of the cocoon is derived by assuming in Equation 5 that the pressure of the cocoon is radiation-dominated, that $t = r/v_{jh}$ (consistently with Murguia-Berthier et al. (2014)), $\Omega = 2\pi[1 - cos(\theta)]$, and where we have used the variable $P'_B = P_{m,B}(1 - \lambda) + \left[(P_{m,B}(1-\lambda))^2 + \frac{4\rho_m L_j v_{jh}}{3\pi r^2}\right]^{1/2}$. Thus, the cocoon opening angle is:

$$\theta_c = acos\left(1 - \frac{P'_B}{4\rho_m v_{jh}^2}\right). \quad (7)$$

The opening angle of the jet is obtained by assuming in Equation 4 that the ram pressure of the jet is that of the jet material that is deflected from its initially radial velocity into a cylindrical flow (for further details, refer to Appendix A). Thus, the jet opening angle is:

$$\theta_j = acos\left(1 - \frac{3P'_B L_j sin^2 \theta_{j,in}}{8c[2L_j \rho_m v_{jh} + 3\lambda \pi r^2 P'_B P_{m,B}]}\right). \quad (8)$$

We must note that the $v_{jh}$, $\theta_c$, and $\theta_j$ were obtained using a Newton-Raphson method. Finally, and consistently with LP19, the energy of the cocoon is:

$$E_c = L_j \left(t - \frac{r}{c}\right). \quad (9)$$

## 3 EFFECTS OF A MAGNETIZED MEDIUM ON THE JET/COCOON DYNAMICS

In this section, we show how the magnetic field and mixing within the cocoon affect the global evolution of the jet and cocoon. We find that the progression and structure of the jet and cocoon are affected when the magnetic field of the medium is high. In addition, we find that the mixing parameter affects the jet and cocoon structure and produces different configurations.

First, it must be noted that in the limit of $B_{0,m} = 0$ G, our model reduces to the model presented by LP19 (Appendix B). Then, to study the magnetization effects in the context of SGRBs, we set the luminosity of the jet equal to $L_j = 2 \times 10^{50}$ erg s$^{-1}$ (consistent with GG23) and the medium akin to the NS-NS GRMHD merger study of (Ciolfi et al. 2017). The density profile of a static medium was set to follow $n = 3$ with a magnitude $\rho_0 = 3.16 \times 10^8$ g cm$^{-3}$ at $r_0 = 1.5 \times 10^7$ cm. The magnetic field profile of the medium was set to have a slope $q = 1.5$ and its magnitude varied between $0 < B_{0,m} < 10^{14.5}$ G (with which its magnetic energy ranged from 0 to 96% of the total energy of the medium). The mixing parameter varied between $0 < \lambda < 1$.

Figure 2 shows the impact of the magnetic field on the $v_{jh}$, $\theta_j$, and $\theta_c$ for two fixed mixing cases: high-mixing case ($\lambda = 0.9$, solid lines) and a non-mixing case ($\lambda = 0$, dashed lines) at three different radii ($10^8$ cm, $10^{8.5}$ cm, and $10^{10}$ cm). Clearly for low magnetic field strengths the jet head velocity, jet and cocoon opening angles, and the energy of the cocoon are unaffected by the magnetic field. On the other hand, for high magnetic field strengths ($B_{0,m} \gtrsim 10^{14}$ G) the jet and cocoon differ from the non-magnetized case. The high-magnetization of the medium will produce a faster jet head (than when no magnetization is present in the medium) for high mixing (and slower for no mixing). A narrower jet, as well as a narrower and less energetic cocoon, will be produced for high mixing, whereas no mixing generates a broader jet and more energetic cocoon (both compared to when the medium has no magnetization). We must note though that in our model the jet in at all times non-magnetized (i.e. there is no magnetic mixing from the cocoon into the jet). The effects of the magnetization of the medium at a larger radius ($> 10^{10}$ cm) are practically negligible.

Figure 3 shows how the mixing parameter affects the evolution of the jet-cocoon through a highly magnetized medium ($B_{0,m} = 2 \times 10^{14}$ G). Note how each variable is normalized over the non-magnetized and control case. This is, $v_{jh}/v_{jh,ctrl}$, $\theta_j/\theta_{j,ctrl}$, $\theta_c/\theta_{c,ctrl}$, and $E_c/E_{c,ctrl}$. The domain covers from $10^8$ cm up to $10^{10}$ cm. The mixing parameter varied between 0 and 1. As a function of the mixing parameter, three clear regimens are present for all the variables. These are the low-mixing regime ($\lambda \lesssim 0.3$), the intermediate-mixing regime ($0.3 \gtrsim \lambda \gtrsim 0.6$), and the high-mixing regime ($\lambda \gtrsim 0.6$). For the low-mixing case, a slower jet head, broader jet, and broader and more energetic cocoon (slowJH-broadJC), compared to when the medium is non-magnetized, will be produced. Meanwhile, for the high-mixing regime a faster jet head, narrower jet, and narrower and less energetic cocoon (fastJH-narrowJC), compared to the non-magnetized case, will be produced. Note that the effects that the magnetic field produces are at most ~ 10% with re-spect to the non-magnetized case. On the other hand, for a chosen jet luminosity, density, and magnetic profiles, the mixing parameter value that separates the slowJH-broadJC from the fastJH-narrowJC is $\lambda \sim 0.3 - 0.4$, except for the cocoon opening angle, whose behavior switches at $\lambda \sim 0.65$. There may be cases in which the faster jet head with a narrow jet may have a broader cocoon opening angle (e.g. at $\lambda = 0.5$). We also note that the $\lambda = 0$ case is different from setting $B_{0,m} = 0$. While a case with $B_{0,m} = 0$ has no magnetic field whatsoever, a case with $\lambda = 0$ does have magnetic field in the ambient, but such field is unable to penetrate in the cocoon region (see Equations 3 and 5).

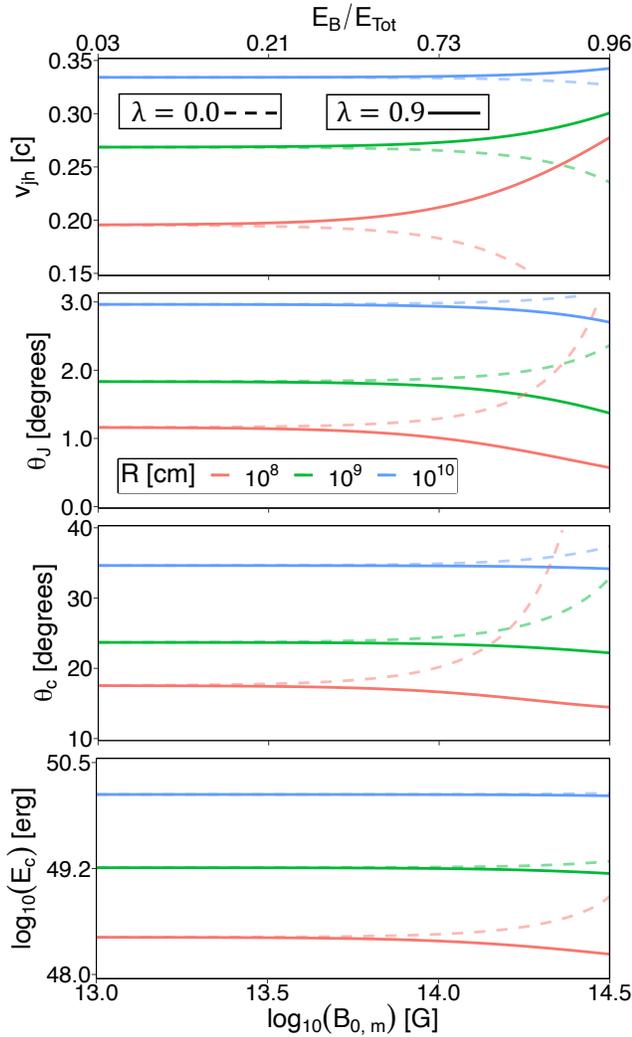

**Figure 2.** Magnetic effects produced on the $v_{jh}$ (upper panel), $\theta_j$ (mid-upper panel), $\theta_c$ (mid-lower panel), and $E_c$ (lower panel) for $\lambda = 0.0$ (dashed lines) and $\lambda = 0.9$ (solid lines). In each panel three different radii are shown ($10^8$ cm in green, $10^9$ cm in red, and $10^{10}$ cm in blue) at different initial magnetic field strengths ($B_{0,m}$, lower axis) or fraction of the magnetic energy ($E_B/E_{Tot}$, upper axis).

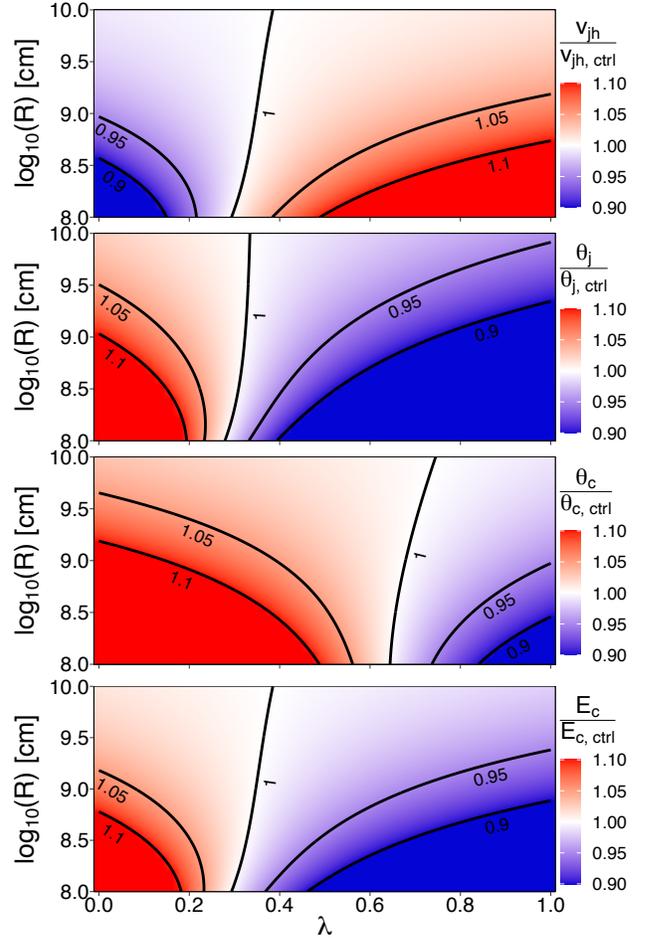

**Figure 3.** Jet-cocoon properties as a function of the mixing parameter $\lambda$ and the radius $R$. From top to bottom, the panels display the jet head velocity $v_{jh}$, the opening angle of the jet $\theta_j$, the opening angle of the cocoon $\theta_c$ and the energy of the cocoon $E_c$. The magnetic field used was $B_{0,m} = 2 \times 10^{14}$ G and all variables are normalized to the non-magnetized model.

## 4 NUMERICAL SIMULATIONS

To validate the model and to constrain the mixing parameter we ran a set of 2D RHD simulations. The simulations were carried out using the RHD code PLUTO (Mignone et al. 2012) in 2D spherical coordinates, and for the simulations to be in accordance with the assumptions of the semi-analytic model, we considered only the magnetic pressure magnitude and not its geometry. The magnetic pressure was therefore simulated by adding additional thermal pressure to the external medium. This approach should be reliable to study the effect of a static ambient medium embedded with a fully tangled[2] magnetic field at small scales.

### 4.1 Setup

The initial setup and the boundary conditions used in the simulations are shown in Figure 4 a. An axisymmetric, relativistic, and collimated jet was injected from the polar axis at a distance $d_{co} = 10^3$ km from the center of the domain. The jet was launched with a constant luminosity $L_j = 2 \times 10^{50}$ erg s$^{-1}$, opening angle $\theta_j = 10°$, and Lorentz factor at infinity $\Gamma_{\infty,j} = 400$. The initial Lorentz value of the jet ($\Gamma_{0,j}$) varied between 8 and 14, depending on the model. The density and magnetic pressure of the medium were based on the simulations of Ciolfi et al. (2017), specifically the equal NS-NS mass model (with soft EOS) in which the delay between the merger and the jet launching was 45 ms and the mass of the ejecta is of order $\sim 10^{-2} M_\odot$. The pressure of the medium was taken to be an ideal gas ($P_{m,g}$ with $T_g = 6 \times 10^7$ K and adiabatic index of 5/3) plus the contribution of the magnetic pressure ($P_{m,B}$). The density and magnetic profiles followed $\rho_m \propto \rho_0 R^{-3}$ and $P_{m,B} \propto B_{0,m}^2 R^{-4}$, where $\rho_0 = 3.16 \times 10^8$ g cm$^{-3}$ and $B_{0,m}$ are the density and the magnetic field strength at $r_0 = 1.5 \times 10^7$ cm. The $B_{0,m}$ varied from between 0 and $10^{14.5}$ G, depending on the model. The axisymmetric and equatorial boundaries were set with reflective conditions. All other boundaries had outflow conditions. The simulations were carried out with a Courant number $Co = 0.3$, and the total integration time was $t = 0.145$ s.

---
[2] We note that incorporating a tangled magnetic field in the initial condition of our two-dimensional simulations was not feasible due to the assumed axis-symmetry.

**Table 1.** Characteristics of the simulations. From left to right: model name, magnetic field ($B_{0,m}$), magnetic energy to total energy ratio ($E_B/E_{Tot}$), and initial Lorentz factor of the jet ($\Gamma_{0,j}$). We note that $E_{Tot}$ spans from $2.6 \times 10^{49}$ erg to $1.9 \times 10^{50}$ erg.

| Model | $B_{0,m}$ [$10^{14}$ G] | $E_B/E_{Tot}$ | $\Gamma_{0,j}$ |
|---|---|---|---|
| B1Γ8 | 1.0 | 0.74 | 8 |
| B1Γ10 | 1.0 | 0.74 | 10 |
| B1Γ12 | 1.0 | 0.74 | 12 |
| B1Γ14 | 1.0 | 0.74 | 14 |
| B0.5Γ10 | 0.5 | 0.41 | 10 |
| B1.25Γ10 | 1.25 | 0.81 | 10 |
| B1.5Γ10 | 1.5 | 0.86 | 10 |
| B2Γ10 | 2.0 | 0.92 | 10 |
| CΓ8 | - | - | 8 |
| CΓ10 | - | - | 10 |
| CΓ12 | - | - | 12 |
| CΓ14 | - | - | 14 |

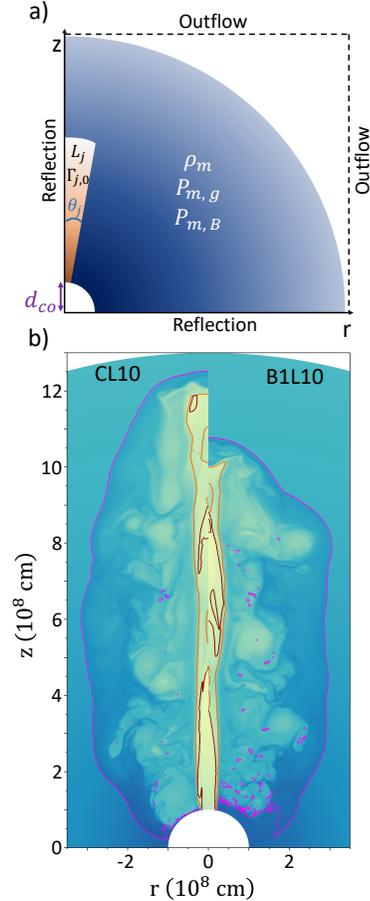

**Figure 4.** a) Numerical setup scheme showing a relativistic and collimated jet evolving through a static medium that had an extra pressure component corresponding to the magnetic pressure. The jet was launched from a distance $d_{co}$, luminosity $L_j$, initial Lorentz factor $\Gamma_{0,j}$, and opening angle $\theta_j$. The medium had a density $\rho_m$, gas pressure $P_{m,g}$, and magnetic pressure $P_{m,B}$. The boundary conditions are also shown. b) Examples of a relativistic jet drilling through two different mediums (left panel: model CΓ10; right panel: model B1Γ10) at t = 0.145 s. The density maps range from $10^3 \text{g cm}^{-3}$ (yellow) to $10^9 \text{g cm}^{-3}$ (navy blue) and the isocontours indicate $\Gamma = 1.0002$ (pink), $\Gamma = 4$ (orange), and $\Gamma = 10$ (brown).

The computational domain extended from $r_{min} = 10^8$ cm to $r_{max} = 1.3 \times 10^9$ cm and from $\theta_{min} = 0$ to $\theta_{max} = \pi/2$. The total mass of the medium in the domain was $8.59 \times 10^{-3}$ M$_\odot$. Consistently with GG23, we used a spherical fixed mesh with $N_r = 10000$ logarithmically radial increasing divisions and $N_\theta = 1000$ equal size angular divisions was used, the finest grid size was $8.2 \times 10^3$ cm.

Eleven models, each one with specific $B_{0,m}$ and $\Gamma_{0,j}$ values were ran. Table 1 shows the details of each simulation. The name of each model indicates the magnetic field strength of the medium ($B_{0,m}$ for magnetized followed by the field strength (in units of $10^{14}$ G) or C for control case), and the $\Gamma_{0,j}$ of the jet.

### 4.2 Mixing parameter constraint

Two examples of the simulations carried out are shown in Figure 4 b. The panel shows density maps and isocontours of Lorentz factors for a magnetized case and the corresponding non-magnetized case (both at $t = 0.145$ s). The left panel shows the case without the magnetic pressure (model CΓ10) while the right panel shows the magnetized case (model B1Γ10). For these models, the jet traveled a shorter distance when evolving through a medium where the magnetic pressure is considered. The latter may or may not be the case for the rest of the models and needs to be quantified.

To quantify the effects that the magnetic pressure produces in the $r_{fs}$, $v_{jh}$, $\theta_j$, and $\theta_c$ the jet and cocoon and medium were distinguished from the each other based on their correspondent $\Gamma_\infty$ (akin to the methodology from Lazzati et al. (2021)). Other methods, which distinguish the jet from the cocoon and from the medium based on $\Gamma$ and on the radial and angular velocity components, have also been proposed (e.g. Hamidani & Ioka 2023a; Hamidani & Ioka 2023b). The external static medium was identified as the material with $\Gamma_\infty = 1$, the cocoon as that with $1 < \Gamma_\infty < 22$, and the jet as that with $\Gamma_\infty \geq 22$. Both $\theta_c$, $\theta_j$ were calculated using $tan(\theta_n) = w/h$, where $h$ is the maximum height and $w$ the maximum width of the jet or cocoon. The $E_c$ was obtained from Equation 9.

The effects of magnetization on the simulations and the analytic model are shown in Figure 5. From top to bottom, the panels show $\theta_c$ (upper-most panel), $\theta_j$ (mid-upper panel), $r_{fs}$ (mid-lower panel), and $E_c$ (bottom panel). Simulation results are shown with circu-lar symbols. Black symbols show the non-magnetized control case, green symbols show the results of tangled field simulations, while the orange symbol shows the result of a poloidal field simulation (from GG23). All simulation results are shown at a time $t = 0.145$ s are shown. Note that every variable is normalized by its non-magnetized case, to highlight the effect of magnetization.

The jets from the simulations had $\Gamma_{0,j} = 10$ and different $B_{0,m}$ (green dots, see Table 1 for details). The analytic solutions for low mixing value ($\lambda = 0.1$, red lines) and high mixing ($\lambda = 0.9$, blue lines), as well as the linear regression best fit for the simulations (black dashed lines), are also shown. For the analytic solutions, we used the same values for all the variables as those in the simulations and set $t = 0.145$ s in equations 6 and 8. Additionally, the results from the 2D RMHD study of GG23 of a jet with $\Gamma_{0,j} = 5$ evolving through a medium with poloidal magnetic field (with $B_{0,m} = 5 \times 10^{12}$ G) at $t = 0.11$ s are shown (orange dot). While our model does not include a directional magnetic field this comparison allows us to evaluate the effect of the geometry of the field in the results.

The magnetic field dependence of the simulations results shows a general trend where the jet and cocoon opening angles and the cocoon energy increase with higher magnetization (which compared to the correspondent non-magnetized case may differ by a factor of $\sim 2$ and $\sim 20\%$, respectively). On the other hand, the forward shock radius decreases with higher magnetization of the medium (up to $\sim 15\%$ compared to the non-magnetized). Finally, the effect of the magnetic field on the cocoon energy can be up to $\sim 10\%$ larger than the non-magnetized case. These dependencies on the magnetic field intensity are better reproduced by a low mixing model ($\lambda \sim 0.1$). Interestingly, the simulation with the poloidal magnetic field of GG23 shows good agreement with our simulations and analytical model. It should be noted that magnetic field strength spans only a factor of two, therefore, the observed effect is limited. A large range of magnetic field strengths should be covered in future works.

To verify that the consistency between the semi-analytical model and the simulations was independent of the Lorentz factor of the jet, we ran an extra set of simulations with varying $\Gamma_{0,j}$ (see Table 1 for details). Figure 6 shows the $\theta_c$ and $\theta_j$ (upper panel), $r_{fs}$ (middle panel) and $E_c$ (bottom panel) for simulations with different $\Gamma_{0,j}$ and $B_{0,m}$ at $t = 0.145$ s. For each Lorentz factor, we simulated the case with $B_{0,m} = 10^{14}$ G as well as its correspondent control case ($B_{0,m} = 0$ G). Also, the simulations shown in Figure 5 are also presented. It is clear that, independently of the chosen $\Gamma_{0,j}$, when moving in a magnetized static medium then the jet tends to move slower while the cocoon is broader and more energetic. No clear trend was found for the jet opening angle.

## 5 SUMMARY AND CONCLUSIONS

In this paper, we study the effects that a static and magnetized medium (with a tangled field) produces on relativistic and non-magnetized jets and their cocoons. We develop as a first approximation a semi-analytical model that includes the effect of the magnetic field present in the medium and is based on the pressure equilibria between the jet head and the medium, between the jet and its cocoon, and between the cocoon and the medium. The magnetization of the medium and the fraction that is entrained into the cocoon are included.

We found that the dynamics of the jet and cocoon could be affected by a highly magnetized medium and by the entrained material. Three regimes were found depending on the fraction of entrained material in the cocoon. For low-mixing a slower-broader jet with a broader and more energetic cocoon would be produced. High-mixing would instead produce a faster-narrower jet with a narrow and less-energetic cocoon. Such effects of the magnetic field can reach a 10% variation. Interestingly, there would be a case in which the faster and narrower jet produces a broader cocoon at intermediate mixing. Additionally, for the case where the magnetic pressure is set to zero, we reproduce the solution found by LP19.

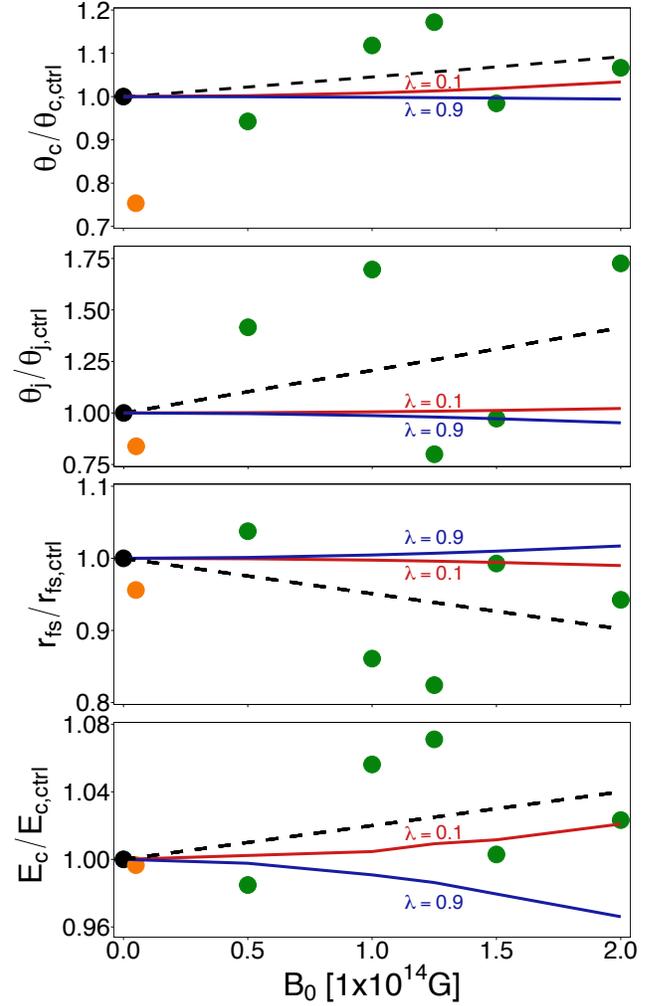

**Figure 5.** Cocoon opening angle ($\theta_c$, upper-most panel), jet opening angle ($\theta_j$, mid-upper panel), forward shock radius ($r_{fs}$, mid-lower panel), and cocoon energy ($E_c$, bottom panel) for different simulations and semi-analytic models as a function of the magnetization of the medium. Note how every variable is normalized by its non-magnetized case. The green dots represent the simulations with $\Gamma_{0,j} = 10$ and different $B_{0,m}$. The black dots represent the correspondent control case simulations without magnetization. The integration time shown is $t = 0.145$ s. The black dashed line represents the linear regression best fit for the simulations. The orange dot represents the 2D RMHD model of GG23 at $t = 0.11$ s. The solid lines represent two different semi-analytical models ($\lambda = 0.1$, red line; $\lambda = 0.9$, blue line).

To validate the semi-analytical model and to constrain the mixing parameter, we ran a set of 2D RHD simulations of a relativistic non-magnetized jet evolving through a static medium in which the magnetic component was simulated as an extra pressure component of the medium. Eleven models with different magnetization and jet Lorentz factors were calculated. We stress that the explored magnetic field parameter space is narrow (factor $\sim 2$) and a wider range of numerical problem. Additionally, the observed effects could not be phenomenological due to our approximations or technical difficulties in measuring different jet parameters, therefore, more work is needed in the future.

We must note that we set the jet and medium properties consistent with those of standard SGRBs. Jets characterized by lower luminosities or wider launching angles will experience reduced ram pressure, leading to a greater prominence of magnetic effects compared to the findings presented in our study. Further studies are required to study the magnetic effects in less powerful jets.

## ACKNOWLEDGEMENTS

We thank the referee for the helpful comments and improvement of the manuscript. LGG and DLC acknowledge the support from the Miztli-UNAM supercomputer (project LANCAD-UNAM-DGTIC-321) for the assigned computational time in which the simulations were performed and also thank the support of the UNAM-PAPIIT grant IG100820. DLC is supported by the Investigadoras e Investigadores por México, CONAHCyT at the Instituto de Astronomía, UNAM. DL acknowledges support from NSF grant AST-1907955. LGG acknowledges support from the CONACyT doctoral scholarship. Many of the images in this study were produced using VisIt. VisIt is supported by the Department of Energy with funding from the Advanced Simulation and Computing Program, the Scientific Discovery through Advanced Computing Program, and the Exascale Computing Project.

## DATA AVAILABILITY

The data underlying this article will be shared on reasonable request to the corresponding author.

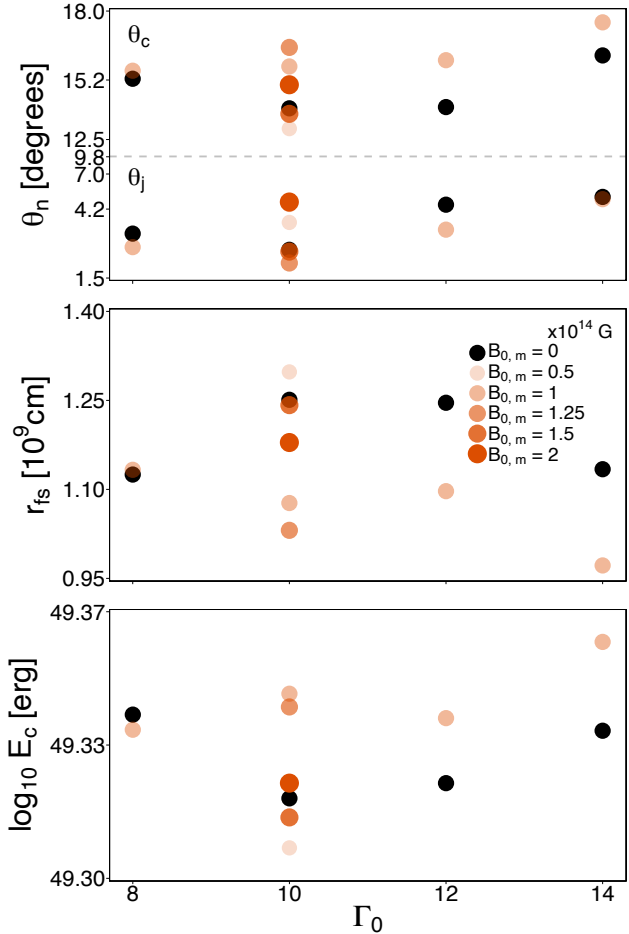

**Figure 6.** Cocoon opening angle $\theta_c$ and jet opening angle $\theta_j$ (upper panel), forward shock radius $r_{fs}$ (middle panel) and cocoon energy $E_c$ (bottom panel) for simulations with different $\Gamma_{0,j}$ and $B_{0,m}$ at $t = 0.145$ s. The size and color of the dots indicate the magnitude of $B_{0,m}$. The black dots represent the correspondent control case without magnetization.

magnetic field strengths to more comprehensively understand the effects of the magnetic field is required. While a constant offset was observed between the analytical and numerical results, in general, the jet head tends to be slower while the opening angle is fairly unaffected by magnetization. Cocoons in magnetized media are inclined to be broader and more energetic. Overall, although the data dispersion, these results were found to be better consistent with the low mixing case of our semi-analytical model. We also carried out a comparison with the simulations of García-García, López-Cámara, & Lazzati (2023) that consider a poloidal magnetic field in the ambient material. We find that their results could also be comparable with the model predictions, showing that field geometry may not be a fundamental factor in predicting outcome.

Our model provides a general frame to study the properties of relativistic jets in different magnetized media. The application could be extended to examine diverse parameters such as jet luminosity, medium density, and magnetic profiles. We stress that our model assumes tangled magnetic fields in the medium, a geometrically ordered magnetic field could alter our results, for example by boosting the jet head velocity, however, by considering different geometry the complexity of the problem increases exponentially turning into a

## REFERENCES

Abbott, B. P., Abbott, R., Abbott, T. D., et al. 2017 A, ApJ, 848, L12. doi:10.3847/2041-8213/aa91c9
Abbott, B. P., Abbott, R., Abbott, T. D., et al. 2017, Phys. Rev. Lett., 119, 161101. doi:10.1103/PhysRevLett.119.161101
Aguilera-Miret R., Viganò D., Carrasco F., Miñano B., Palenzuela C., 2020, PhRvD, 102, 103006. doi:10.1103/PhysRevD.102.103006
Begelman M. C., Cioffi D. F., 1989, ApJL, 345, L21. doi:10.1086/185542
Berger, E. 2014, ARA&A, 52, 43. doi:10.1146/annurev-astro-081913-035926
Bromberg O., Nakar E., Piran T., Sari R., 2011, ApJ, 740, 100. doi:10.1088/0004-637X/740/2/100
Ciolfi R., Kastaun W., Giacomazzo B., Endrizzi A., Siegel D. M., Perna R., 2017, PhRvD, 95, 063016. doi:10.1103/PhysRevD.95.063016
Combi L., Siegel D. M., 2023, ApJ, 944, 28. doi:10.3847/1538-4357/acac29
Costa E., Frontera F., Heise J., Feroci M., in't Zand J., Fiore F., Cinti M. N., et al., 1997, Natur, 387, 783. doi:10.1038/42885
Cowperthwaite P. S., Berger E., Villar V. A., Metzger B. D., Nicholl M., Chornock R., Blanchard P. K., et al., 2017, ApJL, 848, L17. doi:10.3847/2041-8213/aa8fc7
Drout M. R., Piro A. L., Shappee B. J., Kilpatrick C. D., Simon J. D., Contreras C., Coulter D. A., et al., 2017, Sci, 358, 1570. doi:10.1126/science.aaq0049
De Colle, F., Kumar, P., & Aguilera-Dena, D. R. 2018, ApJ, 863, 32. doi:10.3847/1538-4357/aad04d
De Colle, F., Kumar, P., & Hoeflich, P. 2022, MNRAS, 512, 3627. doi:10.1093/mnras/stac742
De Colle, F., Lu, W., Kumar, P., et al. 2018, MNRAS, 478, 4553. doi:10.1093/mnras/sty1282

## APPENDIX A: OBTAINMENT OF THE JET HEAD VELOCITY, JET OPENING ANGLE, AND COCOON OPENING ANGLE

In this Appendix we detail the procedure in order to obtain the solutions for the velocity of the jet head (Equation 6 in the manuscript and Equation A9 in this Appendix), the jet opening angle (Equation 8 and Equation A16), and the cocoon opening angle (Equation 7 and Equation A13).

Let us remind the reader the three pressure balances in which we base our semi-analytical model (equations 3, 4, and 5). These are:

$$P_{j,ram} = P_{mj,ram} + P_{m,B}, \tag{A1}$$

$$P_j = P_c + \lambda P_{m,B}, \tag{A2}$$

$$P_c + \lambda P_{m,B} = P_{mc,ram} + P_{m,B}. \tag{A3}$$

First, the procedure to obtain the velocity of the jet head is next presented. For the latter, we use Equation A1 (i.e. the Rankine-Hugoniot jump condition for momentum conservation), this is:

$$\rho_j h_j \Gamma_j^2 v_j^2 + P_j = \rho_m h_m v_m^2 + P_m + P_{m,B}, \tag{A4}$$

where $\rho_j$ is the jet density, $h_j$ is the jet enthalpy density, $v_j$ is the jet velocity, $\Gamma_j$ is the Lorentz factor of the jet, $P_j$ the jet pressure, $\rho_m$ the medium density, $h_m$ the enthalpy density of the medium, $v_m$ the medium velocity, $P_m$ medium pressure and, $P_{m,B}$ the magnetic pressure of the medium.

Changing the reference system from the inertial frame to that which moves with the working surface of the jet, the Equation A4 becomes:

$$\rho_j h_j \Gamma_j^2 (v_j - v_{jh})^2 + P_j = \rho_m h_m v_{jh}^2 + P_m + P_{m,B}, \tag{A5}$$

where $v_{jh}$ is the jet head velocity.

For a strong shock, where the ram pressure of the jet and medium are far greater than their internal pressure component, and for the case in which the magnetic pressure is not negligible, then Equation A5 reduces to:

$$\rho_j h_j \Gamma_j^2 (v_j - v_{jh})^2 = \rho_m h_m v_{jh}^2 + P_{m,B}, \quad (A6)$$

where the enthalpy of the medium is given by $h_m = 1 + 4(P_m + P_{m,B})/\rho_m c^2$. Since $\rho_m c^2 \gg 4(P_m + P_{m,B})$, then $h_m = 1$ and rearranging terms in Equation A6 gives:

$$\left(1 - \frac{\rho_m}{\rho_j h_j \Gamma_j^2}\right) v_{jh}^2 - 2 v_{jh} v_j + v_j^2 - \frac{P_{m,B}}{\rho_j h_j \Gamma_j^2} = 0. \quad (A7)$$

Equation A7 now has the form of a quadratic equation where $a = 1 - \frac{\rho_m}{\rho_j h_j \Gamma_j^2}$, $b = -2v_j$, $c = v_j^2 - \frac{P_{m,B}}{\rho_j h_j \Gamma_j^2}$, and $x = v_{jh}$. Applying the quadratic formula to A7 we obtain:

$$v_{jh} = v_j \left\{ \frac{1 - \left[1 - \left(1 - \frac{\rho_m}{\rho_j h_j \Gamma_j^2}\right)\left(1 - \frac{P_{m,B}}{\rho_j h_j \Gamma_j^2 v_j^2}\right)\right]^{1/2}}{1 - \frac{\rho_m}{\rho_j h_j \Gamma_j^2}} \right\}. \quad (A8)$$

The enthalpy density of the jet ($h_j = 1 + 4P_j/\rho c^2$) is dominated by the jet pressure, therefore $h_j = 4P_j/\rho c^2$. Then, using $L_j = 4P_j \Gamma_j^2 r^2 \Omega_j c$ and assuming that the jet is ultra-relativistic ($v_j \sim c$) in Equation A8 gives the solution presented in Equation 6 for the jet head velocity:

$$v_{jh} = c \left\{ \frac{1 - \left[1 - \left(1 - \frac{\rho_m c^3 r^2 \Omega_j}{L_j}\right)\left(1 - \frac{P_{m,B} r^2 \Omega_j c}{L_j}\right)\right]^{1/2}}{1 - \frac{\rho_m c^3 r^2 \Omega_j}{L_j}} \right\}. \quad (A9)$$

Next, the procedure to obtain the opening angles of the jet and cocoon (Equations 8 and 7) is presented. For this case, we use the pressure balance which equates the balance at the transition between the cocoon and medium, where the $P_c$ and fraction ($\lambda$) of the entrained magnetic pressure balance with the ram pressure of the medium on the cocoon ($P_{mc,ram}$) plus the $P_{m,B}$. The cocoon's pressure is determined by radiation pressure given by $P_c = E_c/3V_c$, where $E_c = L_j t$ and $V_c = \Omega_c r^3$, thus $P_c = \frac{L_j}{3\Omega_c r^2 v_{jh}}$. Moreover, the time $t$ is related to the jet head velocity and the radius of the jet through $t = r/v_{jh}$ and $P_{mc,ram} = \frac{\rho_m v_{jh}^2 \Omega_c}{\pi}$. Thus, substituting the latter pressures in Equation A3, and reordering terms, one has:

$$-\frac{\rho_m v_{jh}^2}{\pi} \Omega_c^2 - P_{m,B}(1-\lambda)\Omega_c + \frac{L_j}{3r^2 v_{jh}} = 0. \quad (A10)$$

Equation A10 has the form of a quadratic equation and applying the quadratic formula one obtains:

$$\Omega_c = \frac{P_{m,B}(1-\lambda) + \left[(P_{m,B}(1-\lambda))^2 + 4\frac{\rho_m L_j v_{jh}}{3\pi r^2}\right]^{1/2}}{\frac{2\rho_m v_{jh}^2}{\pi}}. \quad (A11)$$

Doing a variable change $P'_B = P_{m,B}(1-\lambda) + \left[(P_{m,B}(1-\lambda))^2 + \frac{4\rho_m L_j v_{jh}}{3\pi r^2}\right]^{1/2}$ and reordering Equation A11 one obtains the solid angle of the cocoon:

$$\Omega_c = \frac{\pi P'_B}{2\rho_m v_{jh}^2}. \quad (A12)$$

Using the relationship between the solid angle and the opening angle, $\Omega = 2\pi[1 - cos(\theta)]$, we obtain the opening angle of the cocoon presented in Equation 7, this is:

$$\theta_c = acos\left(1 - \frac{P'_B}{4\rho_m v_{jh}^2}\right). \quad (A13)$$

Finally, for the solid and opening angles of the jet, we use the pressure balance at the transition between the jet and the cocoon, where the jet internal pressure ($P_j$) is balanced by the cocoon thermal pressure $P_c$ and a fraction ($\lambda$) of the medium's magnetic pressure which has entrained into the cocoon.

Assuming that jet pressure is dominated by its correspondent ram pressure ($P_j = P_{ram,j}\Gamma_j^2$) and using $L_j = 4\Omega_j r^2 c P_{ram,j}\Gamma_j^2$ with a geometrical correction factor $sin^2\theta_{j,in}$ (due to the fact that the jet material is deflected from its initially radial velocity into a cylindrical flow), thus, $P_j = \frac{L_j}{4\Omega_j c r^2} sin^2\theta_{j,in}$. Substituting the jet and cocoon pressures in Equation A2 we obtain the solid angle of the cocoon:

$$\Omega_c = \frac{L_j}{3v_{jh}r^2}\left(\frac{1}{\frac{L_j}{4\Omega_j c r^2} sin^2\theta_{j,in} - \lambda P_{m,B}}\right). \quad (A14)$$

From Equation A12 and Equation A14 we obtain the following solid angle of the jet:

$$\Omega_j = \frac{3\pi P'_B L_j sin^2\theta}{4c[2L_j \rho_m v_{jh} + 3\lambda \pi r^2 P'_B P_{m,B}]}, \quad (A15)$$

from which the opening angle of the jet presented in Equation 8 is:

$$\theta_j = acos\left(1 - \frac{3P'_B L_j sin^2\theta_{j,in}}{8c[2L_j \rho_m v_{jh} + 3\lambda \pi r^2 P'_B P_{m,B}]}\right). \quad (A16)$$

# APPENDIX B: NON-MAGNETIZED CASE AND THE REPRODUCTION OF THE SOLUTIONS FROM LP19

In this Appendix, we show how our solutions for the velocity of the jet head, the jet opening angle, and the cocoon opening angle when the medium is non-magnetized reproduce those from LP19 (for the static medium case).

First, for the jet head velocity when the static ambient medium has $B = 0$ G, we have:

$$v_{jh} = c \left\{ \frac{1 - \left[1 - \left(1 - \frac{\rho_m c^3 r^2 \Omega_j}{L_j}\right)\left(1 - \cancel{\frac{P_{m,B} r^2 \Omega_j c}{L_j}}\right)\right]^{1/2}}{1 - \frac{\rho_m c^3 r^2 \Omega_j}{L_j}} \right\}, \quad (B1)$$

this is:

$$v_{jh} = c \left[ \frac{1 - \sqrt{\frac{\rho_m c^3 r^2 \Omega_j}{L_j}}}{\left(1 - \sqrt{\frac{\rho_m c^3 r^2 \Omega_j}{L_j}}\right)\left(1 + \sqrt{\frac{\rho_m c^3 r^2 \Omega_j}{L_j}}\right)} \right], \quad (B2)$$

where we have changed our notation to that from LP19. Cancelling the numerator and denominator terms we have:

$$v_{jh} = \frac{c}{1 + \sqrt{\frac{\rho_m c^3 r^2 \Omega_j}{L_j}}}. \quad (B3)$$

Meanwhile, the cocoon solid angle for $B = 0$ G, is:

$$\Omega_c = \frac{L_j}{3v_{jh}r^2}\left(\frac{1}{\frac{L_j}{4\Omega_j cr^2}sin^2\theta_{j,in} - \lambda P_{m,B}}\right), \quad (B4)$$

this is:

$$\Omega_c = \frac{4\Omega_j c}{3v_{jh}sin^2\theta_{j,in}}. \quad (B5)$$

For the jet solid angle with the non-magnetized static medium, we have:

$$\Omega_j = \frac{3\pi P'_B L_j sin^2\theta}{4c[2L_j\rho_m v_{jh} + 3\lambda\pi r^2 P'_B P_{m,B}]}, \quad (B6)$$

where now $P'_B = \left[\frac{4\rho_m L_j v_{jh}}{3\pi r^2}\right]^{1/2}$.

Substituting $P'_B$ and eliminating terms, we have:

$$\Omega_j = \frac{1}{4}\sqrt{\frac{3\pi L_j}{\rho_m v_{jh} c^2 r^2}}sin^2\theta, \quad (B7)$$

Finally, we apply the same constraints as those from LP19. The medium was set as a wind ($\rho = \rho_w$) and the time and radius were set to the breakout time and radius of the jet through the ambient medium ($t = t_{bo}$ and $r = r_{bo}$, respectively). Thus, by setting $\rho = \rho_w = \dot{m}_w/\Omega_w v_w$, $t = t_{bo} = v_w \delta t/(v_{jh} - v_w)$ and $r = r_{bo} = v_{jh} t_{bo}$ (where $\dot{m}_w$ is the mass-lose rate, $v_w$ the wind velocity, $\Omega_w$ the wind solid angle, $\delta t$ the time between the merger and the jet launching) in equations B3, B5, and B7 we obtain the same equations as those in LP19 for the static medium case. These are:

$$v_{jh} = \frac{c}{1 + \sqrt{\frac{\dot{m}_w c^3 \Omega_j}{v_w \Omega_w L_j}}}. \quad (B8)$$

$$\Omega_c = \frac{4\Omega_j c v_w \delta t}{3r(v_{jh} - v_w)sin^2\theta_{j,in}}. \quad (B9)$$

$$\Omega_j = \frac{1}{4}\sqrt{\frac{3\pi L_j v_w \Omega_w}{\dot{m}_w v_{jh} c^2}}sin^2\theta, \quad (B10)$$

This paper has been typeset from a T<sub>E</sub>X/LAT<sub>E</sub>X file prepared by the author.